\documentclass[Physsubmission, Phys]{SciPost}
\usepackage{caption}
\usepackage{subcaption}
\usepackage{multicol,caption}
\usepackage{float}
\newcommand{\bbox}{\begin{displaybox}}
\newcommand{\ebox}{\end{displaybox}}

\newcommand{\bea}{\begin{eqnarray}}
\newcommand{\eea}{\end{eqnarray}} 
\usepackage{environ}
\NewEnviron{myequation}{%
\begin{equation}
\scalebox{1.5}{$\BODY$}
\end{equation}
}
\definecolor{orange}{rgb}{1,0.5,0}
\newcommand{\beq}{\begin{displaymath}}
\newcommand{\eeq}{\end{displaymath}}

\newcommand{\E}{{\cal E}}

\newcommand{\pbar}{\overline{\psi}}

\newcommand{\p}{\phi}
\renewcommand{\a}{\alpha}

\newcommand{\tr}{\text{Tr}}

\newcommand{\bx}{\mathbf{x}}
\newcommand{\by}{\mathbf{y}}

\newcommand{\q}{\overline{q}}
\renewcommand{\p}{\psi}
\newcommand{\pb}{\overline{\psi}}
\newcommand{\vx}{\bx}
\newcommand{\vy}{\by}
\newcommand{\vz}{\mathbf{z}}

\newcommand{\T}{{\cal T}}

\newcommand{\n}{\nu}
\newcommand{\m}{\mu}

\newcommand{\oh}{\frac{1}{2}}

\newcommand{\dg}{\dagger}
\newcommand{\non}{\nonumber}
\renewcommand{\t}{\tau}
\newcommand{\rf}[1]{(\ref{#1})}
\newcommand{\ra}{\rightarrow}
\newcommand{\pa}{\partial}

\usepackage{ulem}
\usepackage{xcolor}
\definecolor{olive}{rgb}{0.3, 0.4, .1}
\definecolor{fore}{RGB}{249,242,215}
\definecolor{back}{RGB}{51,51,51}
\definecolor{title}{RGB}{255,0,90}
\definecolor{dgreen}{rgb}{0.,0.5,0.}
\definecolor{gold}{rgb}{1.,0.84,0.}
\definecolor{JungleGreen}{cmyk}{0.99,0,0.52,0}
\definecolor{BlueGreen}{cmyk}{0.85,0,0.33,0}
\definecolor{RawSienna}{cmyk}{0,0.72,1,0.45}
\definecolor{Magenta}{cmyk}{0,1,0,0}

\hypersetup{
    colorlinks,
    linkcolor={red!50!black},
    citecolor={blue!50!black},
    urlcolor={blue!80!black}
}

\usepackage[bitstream-charter]{mathdesign}
\DeclareSymbolFont{usualmathcal}{OMS}{cmsy}{m}{n}
\DeclareSymbolFontAlphabet{\mathcal}{usualmathcal}

\begin{document}

\begin{center}{\Large \textbf{
Excitations of static isolated fermions in the Higgs phase of gauge Higgs theory\\
}}\end{center}

\begin{center}
K.~Matsuyama\textsuperscript{1$\star$},
J.~Greensite \textsuperscript{1}
\end{center}

\begin{center}
{\bf 1} Physics and Astronomy Department, San Francisco State University, San Francisco, USA
* kazuem@sfsu.edu
\end{center}

\begin{center}
\today
\end{center}


\definecolor{palegray}{gray}{0.95}
\begin{center}
\colorbox{palegray}{
  \begin{minipage}{0.95\textwidth}
    \begin{center}
    {\it  XXXIII International (ONLINE) Workshop on High Energy Physics \\“Hard Problems of Hadron Physics:  Non-Perturbative QCD \& Related Quests”}\\
    {\it November 8-12, 2021} \\
    \doi{10.21468/SciPostPhysProc.?}\\
    \end{center}
  \end{minipage}
}
\end{center}

\section*{Abstract}
{
A spectrum of localized excitations of isolated static fermions has been discovered in several different gauge Higgs theories. In lattice numerical simulations, we show that the charged elementary particles can have the spectrum of excitations in the Higgs phase of SU(3) gauge Higgs theory, $q=2$ Abelian Higgs theory, Landau-Ginzburg theory, and in chiral U(1) gauge Higgs theory. Possibly these excited states of the isolated fermions can be observed in ARPES studies of conventional superconductors. Also, we consider that similar kinds of excitations could exist in other gauge Higgs theories, such as the electroweak sector of the Standard Model.}

\vspace{10pt}
\noindent\rule{\textwidth}{1pt}
\tableofcontents\thispagestyle{fancy}
\noindent\rule{\textwidth}{1pt}
\vspace{10pt}

\section{Introduction}
\label{sec:intro}
Molecules, atoms, nuclei, hadrons are composite systems having a spectrum of excitations, but what about the charged ``elementary'' particles? Could quarks and leptons have a spectrum of excitations?

A charged particle is accompanied by a surrounding gauge field (and possibly other fields) as a consequence of Gauss's Law.  These surrounding, localized fields could in principle have a spectrum of excitations. If so, those excitations would look like a mass spectrum of the isolated elementary particle.

Obviously, such excitation doesn't happen in pure QED because any energy eigenstate containing a static $\pm$ charge pair is just the
Coulomb field plus some number of photons. But this could be different in the gauge Higgs theories.  
\subsection{Pseudomatter fields}
\label{pseudo}
In connection with gauge theories we often ask: are all physical states gauge invariant? The answer is: not quite. Note that the Gauss law constraint only requires invariance under infinitesimal gauge transformations, but this does not exclude certain global transformations.  
As a simple example taken from QED, consider a single static charge at point $\vx$ in an infinite volume.  The corresponding physical state
of lowest energy, first written down by Dirac \cite{Dirac:1955uv}, is 
\bea
     |\Psi_\vx \rangle = \pbar^+(\vx) \rho_C(\vx;A) |\Psi_0 \rangle ~~~,~~~
     \rho_C(\vx;A) = \exp\left[-i {e\over 4\pi} \int d^3z ~ A_i(\vz) {\pa \over \pa z_i}  {1\over |\vx-\vz|}  \right]\ .
 \label{rho}
\eea
The state $ |\Psi_\vx \rangle$ satisfies the Gauss Law.  However, considering an arbitrary U(1) gauge transformation, $g(x) = e^{i\theta(x)}$, we separate out the zero mode $\theta(x) = \theta_0 + \tilde{\theta}(x)$.
Then this transforms the static charge operator as $\psi(\vx) \ra e^{i\theta(\vx)}\psi(\vx)$, but the $\rho_C$ operator in Eq.\ \rf{rho} transforms without the zeroth mode $\rho_C(\vx;A) \ra e^{i\tilde{\theta}(x)} \rho_C(\vx;A)$. Then the operator combining the static charge operator and the $\rho_C$ operator together transforms as $|\Psi_\vx \rangle \ra e^{-i\theta_0} |\Psi_\vx \rangle$, so $|\Psi_\vx \rangle$ transforms under the global subgroup of the gauge group. This result reminds us that while Elitzur's theorem says that local symmetries cannot break spontaneously, global symmetries can.

We call operators like $\rho_C$ in Eq.\ \rf{rho}  ``pseudomatter'' fields  \cite{2017_greensite}.  These are non-local functionals of the gauge field which transforms like a matter field in the fundamental representation of the gauge group, except under the global center subgroup of the gauge group. In our work in the gauge Higgs theory, we create physical states by combining the scalar field and pseudomatter fields with the static charge operator.
 
Examples of pseudomatter fields include (i)  Any SU(N) gauge transformation $g_F(\vx;A)$ to a physical gauge $F(A)=0$.  This can be decomposed into $N$ pseudomatter fields $\{\rho_n\}$, and vice-versa, via $ \rho^a_n(\vx;A) = g_F^{\dg an}(\vx;A)$ 
(in fact the operator $\rho_C^*(\vx;A)$ in \rf{rho} is the gauge transformation
to Coulomb gauge in an abelian theory).  And (ii) any eigenstate $\xi_n(\vx;U)$ of the covariant Laplacian operator, $-D^2 \xi_n = \kappa_n \xi_n$, in an SU(N) gauge theory, where
\bea
  (-D^2)^{ab}_{\vx \vy} =  \sum_{k=1}^3 \left[2 \delta^{ab} \delta_{\vx \vy} - U_k^{ab}(\vx) \delta_{\vy,\vx+\hat{k}} 
       - U_k^{\dg ab}(\vx-\hat{k}) \delta_{\vy,\vx-\hat{k}}  \right] \ ,
\label{D2}
\eea
is a pseudomatter field.
 
Pseudomatter fields play an important role in the formulation of excited states of elementary fermions in gauge Higgs theories. 
For static quarks in a pure gauge theory there is a tower of energy eigenstates 
\bea
          \Psi_n(R)  = \q(\vx) V_n(\vx,\vy;U) q(\vy) \Psi_0\ ,
\eea
which we attribute to the string excitations. In fact, these excitations have been observed in computer simulations in \cite{2003_Juge} and in \cite{2016_Brandt}.

A similar spectrum of excitations (metastable due to string breaking) exists in the confinement phase of a gauge Higgs theory. For light quarks, the flux tube forms between the pair of the quark and antiquarks, and the excited hadronic states lie on linear Regge trajectories. But, what about in the Higgs phase? Is there a similar tower of metastable states given by 
\bea
   \Psi_n(R)  = \q^a(\vx) \left[\sum_m c^{(n)}_m \rho_m^a(\vx)  \rho_m^{\dg b}(\vy) \right] q^b(\vy) \Psi_0\ ,
\eea  
where the $\{ \rho_m(\vx) \}$ are pseudo-matter fields? We asked this question in four different models, first in SU(3) gauge Higgs theory \cite{2020_greensite}, then in $q=2$ Abelian gauge Higgs theory \cite{2021_matsuyama}, in Landau-Ginzburg effective action for superconductivity \cite{2021_gm}, and in chiral U(1) gauge Higgs theory (Smit-Swift formulation) \cite{2021_greensite}. In those four models, we impose a unimodular constraint \\ $\phi^*(x) \phi(x)=1$ for simplicity of our calculations. Of course, the four models are different, so each model has its own special features which must be taken into account.
\subsection{Transfer matrix}
Let $E_1(R)$ be the lowest energy, above the vacuum energy $\E_0$, of all states containing a static fermion-antifermion pair separated
by distance $R$, and let  $|\Psi(R)\rangle$ be some arbitrary state of this kind. Then on general grounds
\bea
\langle \Psi(R) | \T^T |\Psi(R) \rangle &=& \sum_n c_n e^{-E_n(R) T} \ra c_1 e^{-E_1(R) T}  ~~~\mbox{as~~} T\ra \infty\ .
\eea
where $\T = e^{-(H-\E_0)a}$ is the transfer matrix ($\tau=e^{-Ha}$) rescaled by an exponential $e^{\E_0 a}$ of the vacuum energy $\E_0$ (from here on we refer to $\T$, rather than $\t$ as the transfer matrix). But this is not very useful for finding the energy of the excited states, because all you get is the ground state in this way.

Alternatively, we may choose some set of states $\{|\Phi_\a(R)\rangle \}$, spanning a subspace of the full Hilbert space with the two static charges. One could then obtain an approximate mass spectrum by diagonalizing $\T$ in the given subspace, as is done in many lattice QCD calculations. However, this requires using a rather large set containing on the order of hundreds of states. Obviously, this method is also not practical for our purposes, where generating the required pseudomatter operators is a computationally intensive process.

As a practical solution for our purposes, we instead generate a small set of states $\{|\Phi_\a(R)\rangle \}$, diagonalize either the transfer matrix $\T$ or a power of the transfer matrix $\T^p$ in the small subspace spanned by these states, and evolve these states in Euclidean time.  The idea is that one or more of the eigenstates $|\Psi_n\rangle$ may be orthogonal, or nearly orthogonal, to the true ground state.  If $|\Psi\rangle$ is orthogonal to the ground state, then
\bea
\langle \Psi| \T^T |\Psi \rangle &=& \sum_n c_n e^{-E_n(R) T}  
 \ra c_{ex} e^{-E_{ex}(R) T}  ~~~\mbox{at large $T$}\ .
\eea
However this method is also not guaranteed to work, so we just need to try it to see if it works or not. 

\section{Models and Results}
\subsection{SU(3) gauge Higgs theory}
Let $\xi_n$ denote the eigenstates  $ -D^2 \xi_n = \kappa_n \xi_n$ of the lattice Laplacian operator in \rf{D2} in SU(3) gauge Higgs theory. At each quark separation $R=|\vx-\vy|$, we consider the 4-dimensional subspace of the Hilbert space spanned by three quark-pseudomatter states, and one quark-scalar state
\bea
          \Phi_n(R) &=& [\q^a(\vx) \xi_n^a(\vx)] ~\times~  [\xi_n^{\dg b}(\vy) q^b(\vy)] ~\Psi_0 ~~~~ (n=1,2,3)  \non \\
          \Phi_4(R) &=& [\q^a(\vx) \phi^a(\vx)] ~\times ~  [\phi^{\dg b}(\vy) q^b(\vy)] ~ \Psi_0 \ .
\label{basis1}
\eea
For this non-orthogonal basis, we calculate numerically the matrix elements and overlaps,
\bea
          [\T]_{\alpha \beta}(R) &=& \langle \Phi_\alpha | \T | \Phi_\beta \rangle  ~~~,~~~
          \left[O\right]_{\alpha \beta}(R) = \langle \Phi_\alpha | \Phi_\beta \rangle \ . 
\eea   
We obtain the eigenvalues of $\T$ in the subspace by solving the generalized eigenvalue problem,
\bea
         [\T] \vec{\upsilon}_n = \lambda_n [O] \vec{\upsilon}^{(n)} ~~\mbox{and}~~
          |\Psi_n(R)\rangle = \sum_{i=1}^4 \upsilon^{(n)}_i |\Phi_i(R)\rangle\ .
\label{Geig}
\eea
The $|\Psi_n(R)\rangle$ are the linear combinations of the non-orthognal basis states $|\Phi_i(R)\rangle$ , and the set of states
$|\Psi_n(R)\rangle$ are the energy eigenstates (i.e.\ eigenstates of the transfer matrix) of the isolated static pair only in the restricted
subspace. Next, we consider evolving states for Euclidean time $T$, and compute
\bea
         \T^T_{nn}(R) &=& \langle \Psi_n | \T^T | \Psi_n \rangle \label{Tnn} \label{TT}  
                                 =  \upsilon^{(n)*}_i  \langle \Phi_i | \T^T | \Phi_j \rangle \upsilon^{(n)}_j  ~~\mbox{with}~~
           E_n(R,T) = - \log \left[{\T^T_{nn}(R) \over \T^{T-1}_{nn}(R) }\right] \ ,
\eea
where $E_n(R,T)$ is a lattice logarithmic time derivative, and can be understood as the energy expectation value of the state 
 $ \Psi\Bigl(R, \oh(T-1)\Bigr) = \T^{(T-1)/2} \Psi(R)$
 which is obtained by evolving $\Psi(R)$ by $\oh (T-1)$ units of Euclidean time.

In order to compute Eq.\ \rf{TT}, we first integrate out the massive (i.e.\ static) fermion fields, and this generates a pair of Wilson lines. Then the numerical computation of $ \langle \Phi_i | \T^T | \Phi_j \rangle$  boils down to calculating the expectation values of products of Wilson lines each terminated by matter or pseudomatter fields.

 There are three possibilities: (i) $\Psi_n(R)$ is an eigenstate in the full Hilbert space, and $E_n(R) = E(R,T)$ is time independent;
 (ii)  $\Psi_n(R)$ evolves to the ground state, and $E_n(R,T) \ra E_1$; (iii) $\Psi_n(R)$ evolves in Euclidean time to a stable or metastable excited state above the ground state. Then $E_n(R,T)$ converges to a value greater than $E_1$. 
For our numerical work, we have computed $E_n(R,T)$ in SU(3) gauge theory with a unimodular Higgs field on a $14^3 \times 32$ lattice volume, with $\gamma=0.5$ and $\gamma=3.5$, in the confinement and Higgs phases respectively. The action is
\bea
        S= - {\beta \over 3} \sum_{plaq} \mbox{ReTr}[U_\m(x)U_\n(x+\hat{\m})U_\m^\dg(x+\hat{\n}) U^\dg_\n(x)] 
        - \gamma \sum_{x,\m} \mbox{Re}[\phi^\dg(x) U_\m(x) \phi(x+\widehat{\m})]\ .
\eea

\begin{figure}[h]
\centering
\begin{subfigure} [b]{0.4\textwidth}
\centering
 \includegraphics[scale=0.5]{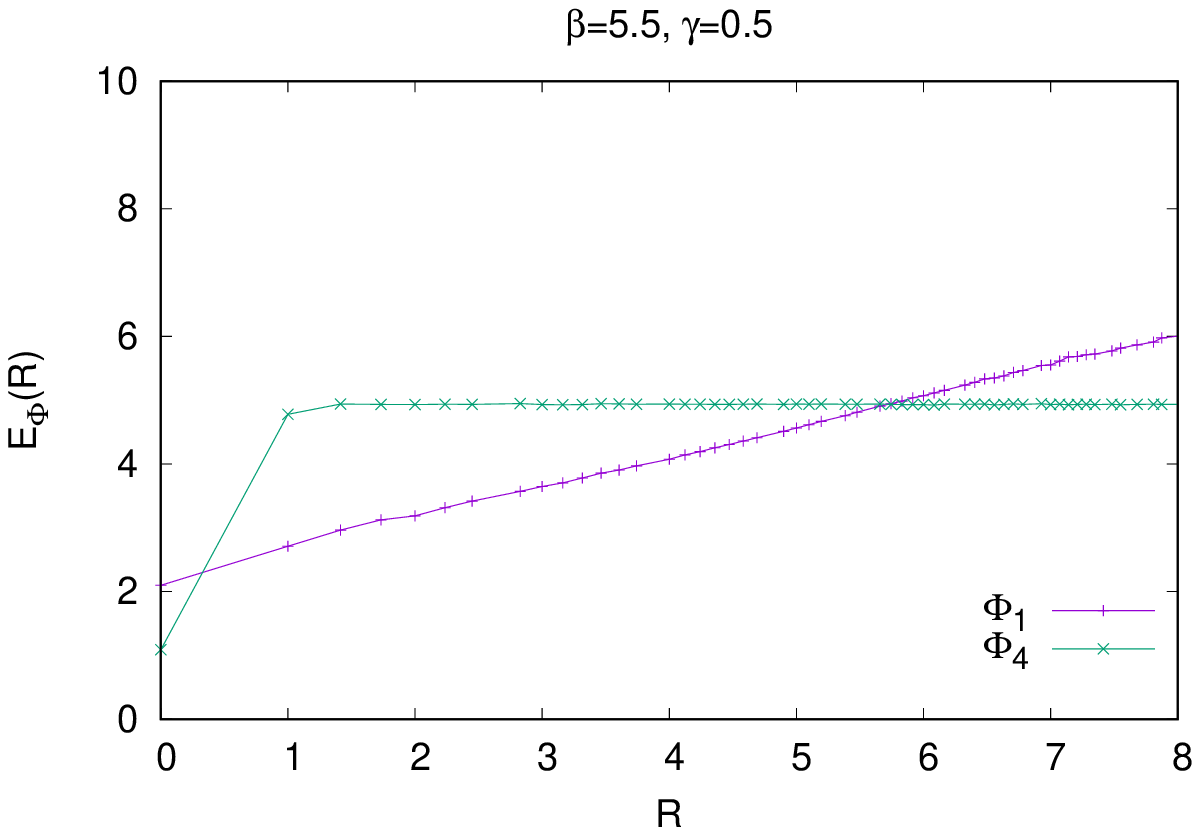}
 \caption{}
 \label{EV05}
 \end{subfigure}
 \hfill
\begin{subfigure} [b]{0.4\textwidth}
\centering
 \includegraphics[scale=0.5]{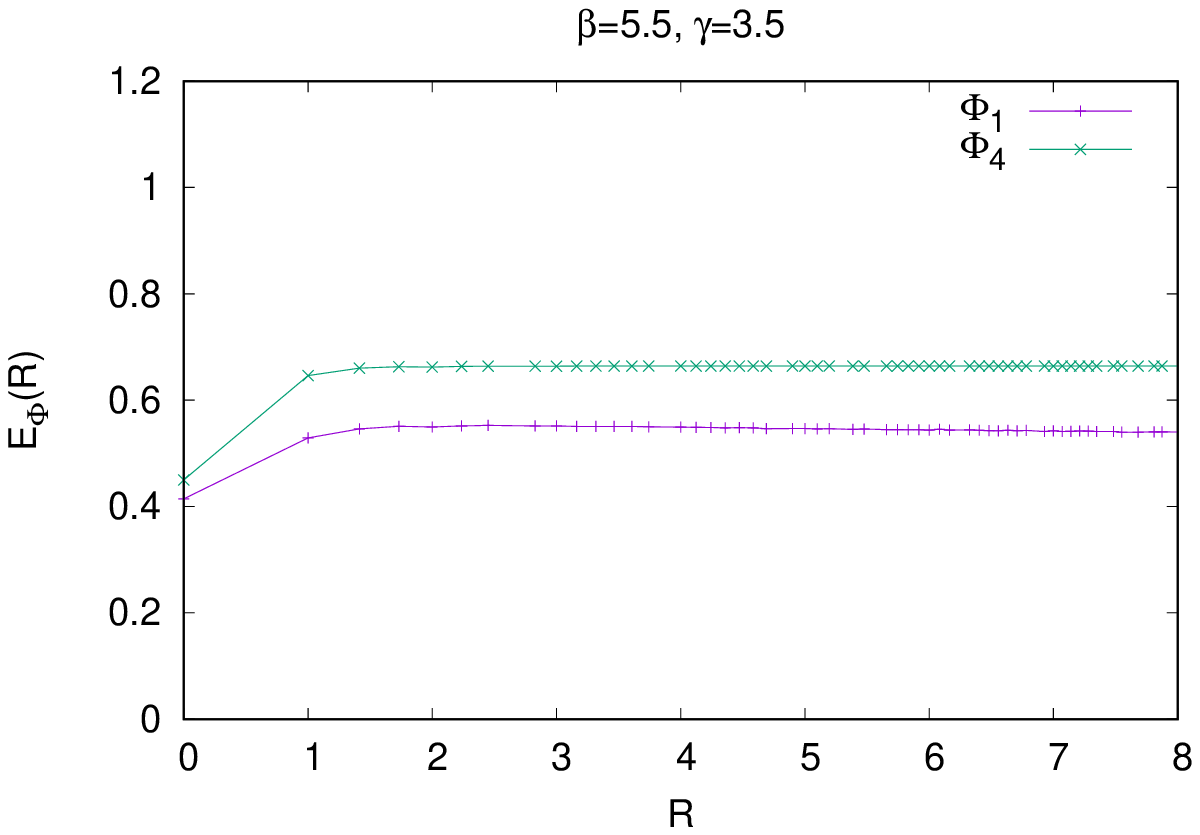}
 \centering
 \caption{}
 \label{EV35}
 \end{subfigure}
 \caption{(a) Energy expectation value of $\Phi_1(R)$ purple line and $\Phi_4(R)$ green line in the confinement phase. (b) Energy expectation value of $\Phi_1(R)$ purple line and $\Phi_4(R)$ green line in the Higgs phase. Figure from \cite{2020_greensite}.}
 \label{fig2}
\end{figure}

Now let us consider two states in particular, 
\bea
          \Phi_1(R) = [\q^a(\vx) \xi_1^a(\vx)] \times [\xi_1^{\dg b}(\vy) q^b(\vy)] \Psi_0 ~~,~~
          \Phi_4(R) = [\q^a(\vx) \phi^a(\vx)] \times  [\phi^{\dg b}(\vy) q^b(\vy)] \Psi_0
\eea
$\Phi_4$ is just a pair of color neutral objects, which can be separated to $R\ra \infty$ with a finite cost in energy. The distinction between the Higgs and confinement phases is that in the confinement phase the energy of every pseudomatter state (such as $\Phi_1$) diverges as $R \ra \infty$, no matter which pseudomatter field is used.
That is the definition of separation-of-charge (S$_c$) confinement \cite{2017_greensite}, which is associated with metastable
flux tubes and Regge trajectories.  S$_c$ confinement disappears in the Higgs phase, where the global center subgroup of the gauge group is spontaneously broken \cite{Greensite:2020nhg}, and this is seen in Fig. \ref{fig2},  with data taken at $\beta=5.5, \gamma=0.5$ in the confinement phase, and $\beta=5.5, \gamma=3.5$ in the Higgs phase. We also find that the overlap $\langle \Phi_1|\Phi_4\rangle \ra 0$
at large $R$ in the confinement phase, but is non-zero in the Higgs phase.

\begin{figure}[H]
\centering
\begin{subfigure} [b]{0.4\textwidth}
\centering
 \includegraphics[scale=0.5]{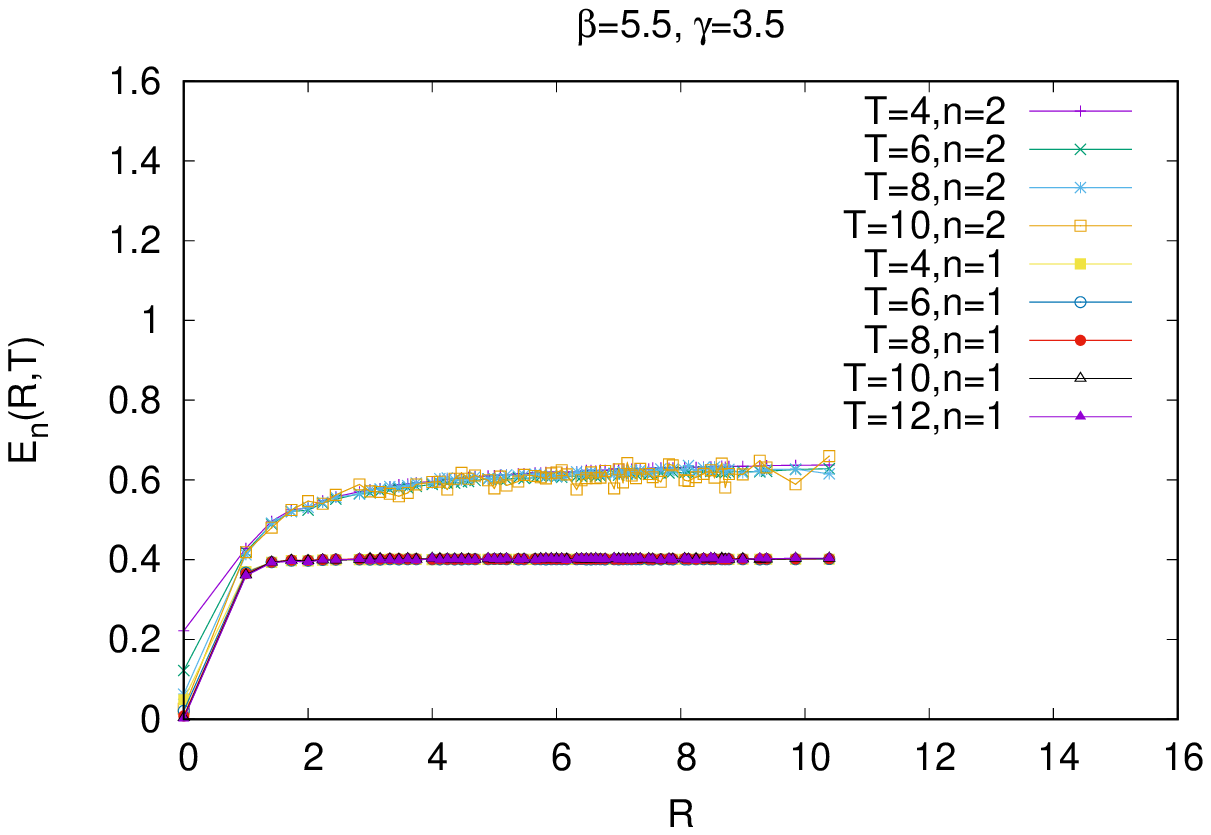}
 \caption{}
 \label{EValt}
 \end{subfigure}
 \hfill
 \begin{subfigure} [b]{0.4\textwidth}
\centering
 \includegraphics[scale=0.5]{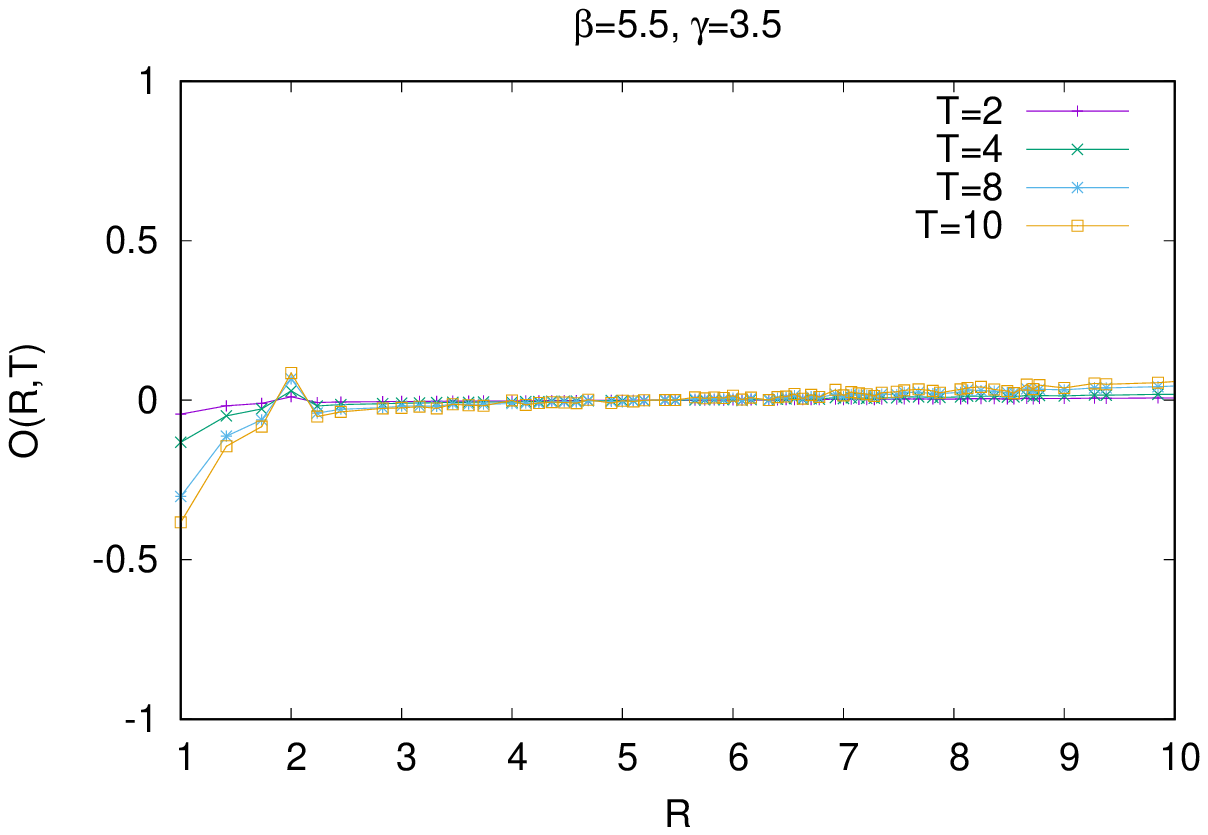}
 \caption{}
 \label{ORT}
 \end{subfigure}
 \caption{(a) energy expectation value $E_n(R,T)$ in the Higgs phase of SU(3) gauge Higgs theory; (b) overlap of $\Psi_1(R), \Psi_2(R)$ after evolution in Euclidean time in the Higgs phase of SU(3) gauge Higgs theory.  Figure from \cite{2020_greensite}.}
 \label{fig3}
 \end{figure}

We solve the generalized eigenvalue problem \rf{Geig} in the non-orthogonal basis \rf{basis1} in the Higgs phase and determine the eigenstates $\Psi_n(R)$ of the pair of static fermion and antifermion. Then we compute the time dependent energy expectation values, $E_n(R,T)$, and the overlap of $\Psi_1(R), \Psi_2(R)$ after evolution for  $T=4-12$ units of Euclidean time.
The results are shown in Fig.\ \ref{fig3}. 
 In Fig.\ \ref{EValt}, time evolution of the energy expectation value of $\Psi_1(R)$, the ground state, converges to the purple line, and the time evolution of the energy expectation value of $\Psi_2(R)$, the first excited state, converges to yellow line, which is the different energy level from the  ground state for $T=4-12$. The energy gap is far smaller than the threshold for vector boson creation. In Fig.\ \ref{ORT}, we see that after some Euclidean time evolution, the ground state $\Psi_1(R)$ and the first excited state $\Psi_2(R)$ are orthogonal to each other. These results in Fig.\ \ref{fig3} are the clear evidence of existence of a stable localized excited state, which is orthogonal to the ground state, in the excitation spectrum of the static fermion and antifermion pair in the Higgs phase of the SU(3) gauge Higgs theory. 
 
\subsection{$q=2$ Abelian Gauge-Higgs theory}
We investigate the localized excited states in $q=2$ Abelian gauge Higgs theory with the action, 
\bea
 S  = - \beta \sum_{plaq}  \mbox{Re}[U_\m(x)U_\n(x+\hat{\m})U_\m^*(x+\hat{\n}) U^*_\n(x)] 
             - \gamma \sum_{x,\m}  \mbox{Re}[\phi^*(x)U^2_\m(x) \phi(x+\widehat{\m})] \ . 
\eea
In this theory, the scalar field has charge $q=2$ as do Cooper pairs. Similarly to SU(3) gauge Higgs theory, we impose a unimodular constraint ${\phi^*(x) \phi(x) = 1}$ for simplicity of our calculations. This is a relativistic generalization of the Landau-Ginzburg effective model of superconductivity.

In our calculation we make use of the four lowest-lying Laplacian eigenstates $\xi_i$ and the Higgs field, defining
$\zeta_i(x)=\xi_i(x), ~i=1-4$ and $\zeta_5(x)=\phi(x)$.
 We define 
 \bea
  Q_\a(R) &=& \pb(\vx) V_\a(\vx,\vy;U) \p(\vy) ~~\mbox{and}~~ V_\a(\vx,\vy;U) =   \zeta_\a(\vx;U) \zeta^*_\a(\vy;U)   \ ,
 \eea 
and also
\bea
  [\mathcal{T}]_{\alpha\beta} = \langle \Phi_\alpha | e^{-(H-\mathcal{E}_0)} | \Phi_\beta \rangle  = \langle Q_\alpha^\dagger(R,1) Q_\beta(R,0) \rangle ~,~
     \left[ O \right]_{\alpha\beta} =  \langle \Phi_\alpha | \Phi_\beta \rangle  = \langle Q_\alpha^\dg(R,0) Q_\beta(R,0) \rangle
\eea 
obtaining the five orthogonal eigenstates of $\mathcal{[T]}_{\alpha\beta}$  by solving the generalized eigenvalue problem  \rf{Geig}, with eigenvalues $\lambda_n$  ordered such that $\lambda_n$ decreases with $n$. Then we consider evolving the states $\Psi_n$ in Euclidean time,
\bea
         \mathcal{T}_{nn}(R,T) &=& \langle \Psi_n | \T^T | \Psi_n \rangle \label{Tnn}  
                                 = \upsilon^{*(n)}_\alpha  \langle Q_\alpha^\dagger(R,T) Q_\beta(R,0) \rangle \upsilon^{(n)}_\beta  \ , 
 \eea
where Latin indices indicate matrix elements with respect to the $\Psi_n$ rather than the $\Phi_\alpha$, and there is a sum
over repeated Greek indices.
After integrating out the massive fermions, whose worldlines lie along timelike Wilson lines (denoted $P(\vx,t,T)$ which are products of squared timelike link variables $U_0^2$ (because charge $q=2$)), we have
\bea
       \langle Q_\alpha^\dagger(R,T) Q_\beta(R,0) \rangle
       = \langle \tr[V^\dagger_\alpha(\vx,\vy;U(t+T)) P^\dagger(\vx,t,T) V_\beta(\vx,\vy;U(t)) P(\vy,t,T)]  \rangle \ ,
 \label{qdq}
\eea
and then use \rf{qdq} to compute the time dependent matrix elements of the transfer matrix as in Eq.\ \rf{Tnn} numerically.
On general grounds, $\mathcal{T}_{nn}(R,T)$ is a sum of exponentials
\bea
      \mathcal{T}_{nn}(R,T) ~=~ \langle \Psi_n(R) | e^{-(H-\mathcal{E}_0)T} | \Psi_n(R) = \sum_j |c^{(n)}_j(R)|^2 e^{-E_j(R) T} \ , 
\eea
where $c_j^{(n)}(R)$ is the overlap of state $\Psi_n(R)$ with the j-th energy eigenstate of the Abelian Higgs theory containing a static fermion-antifermion pair at separation $R$, and $E_j(R)$ is the corresponding energy eigenvalue minus the vacuum energy. 

\begin{figure}[H]
\begin{subfigure} [b]{0.4\textwidth}
\centering
\includegraphics[scale=0.5]{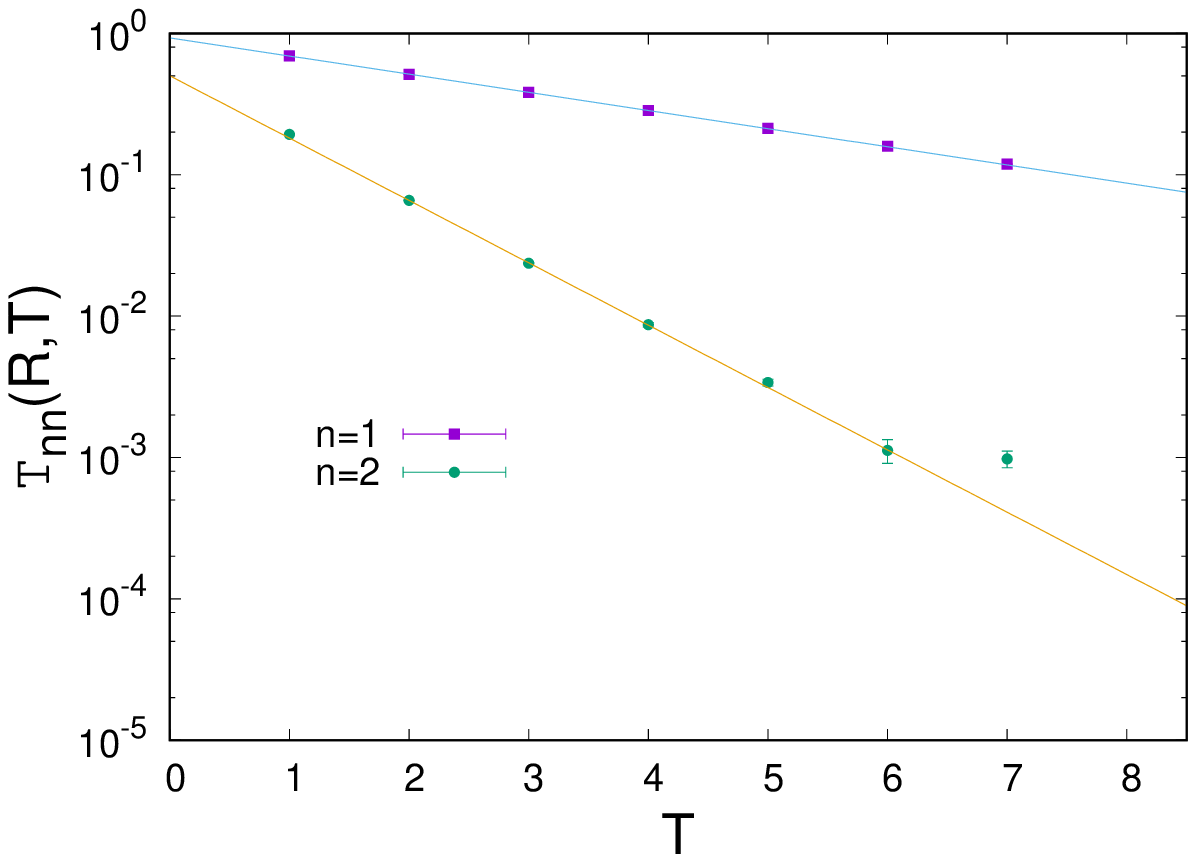}   
\caption{}
\label{fit12}
\end{subfigure}
\hfill
\begin{subfigure} [b]{0.4\textwidth}
\centering
\includegraphics[scale=0.5]{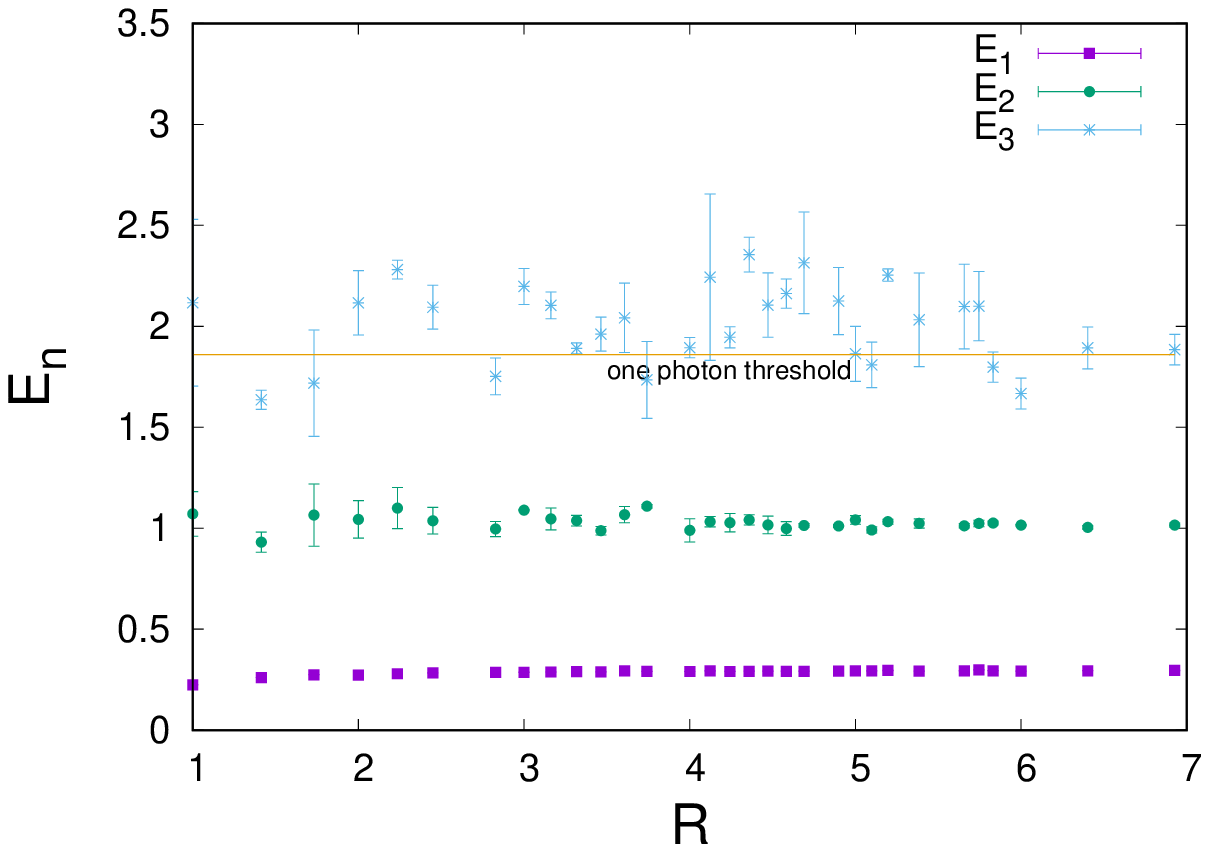}
\caption{}
\label{ER12}
\end{subfigure}   
\caption{(a) An exponential fitting example at $R=6.93$ on a $16^4$ lattice with couplings $\beta=3, \gamma=0.5$. (b) A  plot of the energy expectation values $E_n(R)$ vs.\ $R$ for $n=1,2,3$. Figure from \cite{2021_matsuyama}.}
\label{fig4}
\end{figure}

For our numerical study, we investigate the Higgs region at $\beta$=3 and $\gamma$=0.5. We compute the photon mass from the plaquette-plaquette correlator to be 1.57 in lattice units. The energies $E_n(R)$ for $n=1,2$ are also obtained by fitting the data for $\mathcal{T}_{nn}(R,T)$ vs.\ $T$,  at each $R$, to an exponential falloff. An example of these fits at $R=6.93$ on a $16^4$ lattice with couplings $\beta=3, \gamma=0.5$ are shown in Fig. \ref{fit12}. Fitting through the points at $T=2-5$, we find $E_1=0.2929(6)$ and $E_2(R) = 1.01(1)$. 
We repeated the single exponential fitting analysis for each separation distance $R$;  the data and errors were obtained from ten independent runs, each of 77,000 sweeps after thermalization, with data taken every 100 sweeps, computing $\mathcal{T}_{nn}$ from each independent run.

We also looked for any indication of a second stable excited state by fitting  $\mathcal{T}_{33}(R,T)$ to a sum of exponentials, but of course such an analysis must be treated with caution.  With this caveat, all values of $E_1, E_2, E_3$ together with the one photon threshold are shown in Fig.\ \ref{ER12}. The yellow line is the one photon threshold energy line which is simply $E_1 + m_{photon} = 1.86(1)$ in lattice units. The most important observation is that $E_2(R)$ lies well below this threshold, which implies that the first excited state of the static fermion-antifermion pair is stable. The second excited state $E_3(R)$ seems to lie above or near the one photon threshold is probably a combination of the ground state plus a massive photon.

\subsection{Effective Landau-Ginzburg model}
 The effective Landau-Ginzburg model for ordinary superconductivity is a non-relativistic $q=2$ Abelian Higgs model of this form:
\bea
S = -\beta \sum_{plaq} \text{Re}[UUU^*U^*] - \gamma  \sum_x  \sum_{k=1}^3 \phi^*(x)U^2_k(x)\phi(x+\hat{k})  
   - {\gamma \over \upsilon^2} \sum_x \phi^*(x)U^2_0(x)\phi(x+\hat{t})\ ,
\eea 
where $\upsilon \sim 10^{-2}$ in natural units, is on the order of the Fermi velocity in a metal, and \newline ${\beta={1\over e^2}=10.9}$, where $e$ is the electric charge. In the simulations we go to unitary gauge, where $U_0(x) \approx \pm 1$.  The aim is to find excitations
around pairs  of static $q=\pm 1$ (e) charges, having in mind electrons and holes. 

Couplings  $\gamma, \beta$ determine the photon mass, which is the inverse to the penetration depth, in lattice units.  Therefore  the penetration depth, at given $\gamma$, sets the lattice spacing in physical units.  Unfortunately in this case we found that eigenstates of $\T$ in the subspace have energies which flow, in Euclidean time, to the ground state energy.

\begin{figure}[H]
\begin{subfigure} [b]{0.4\textwidth}
\centering
\includegraphics[scale=0.5]{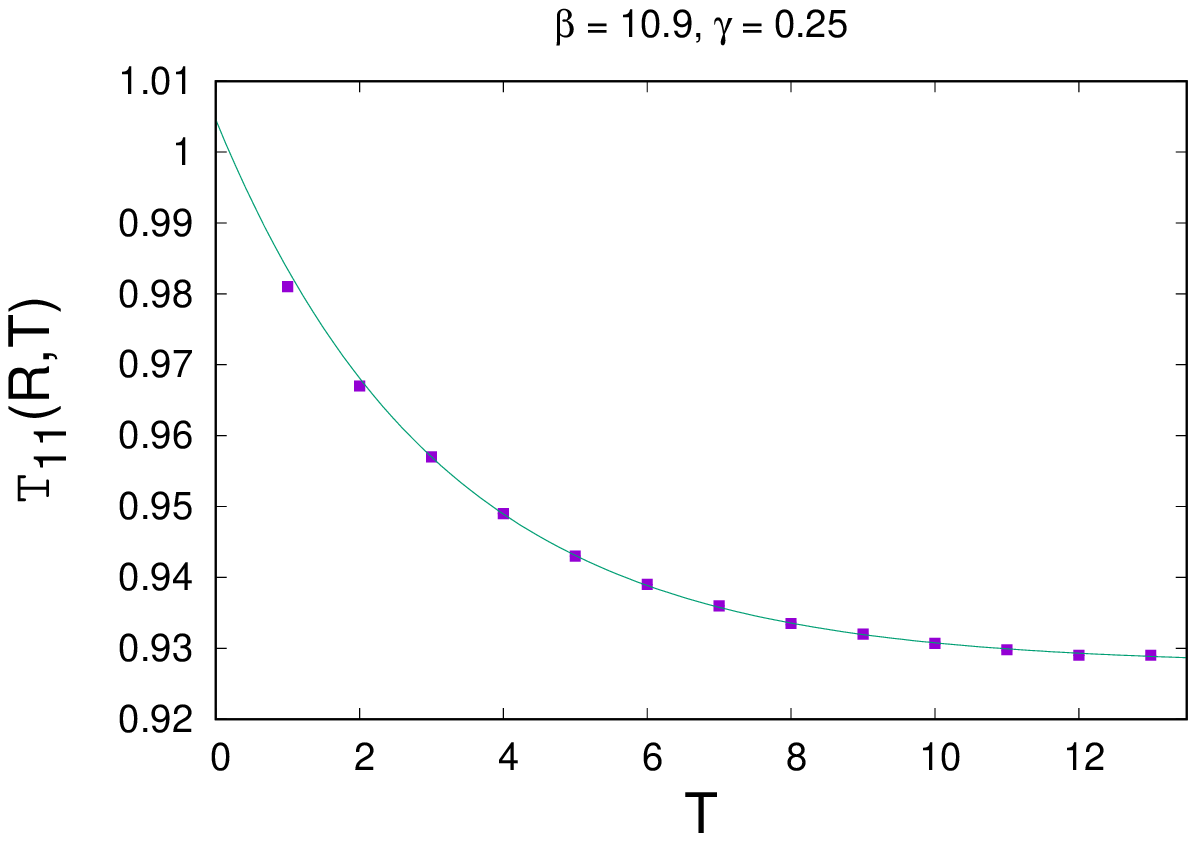} 
\caption{}
\label{fig7a}
\end{subfigure}
\hfill
\begin{subfigure} [b]{0.4\textwidth}
\centering
\includegraphics[scale=0.5]{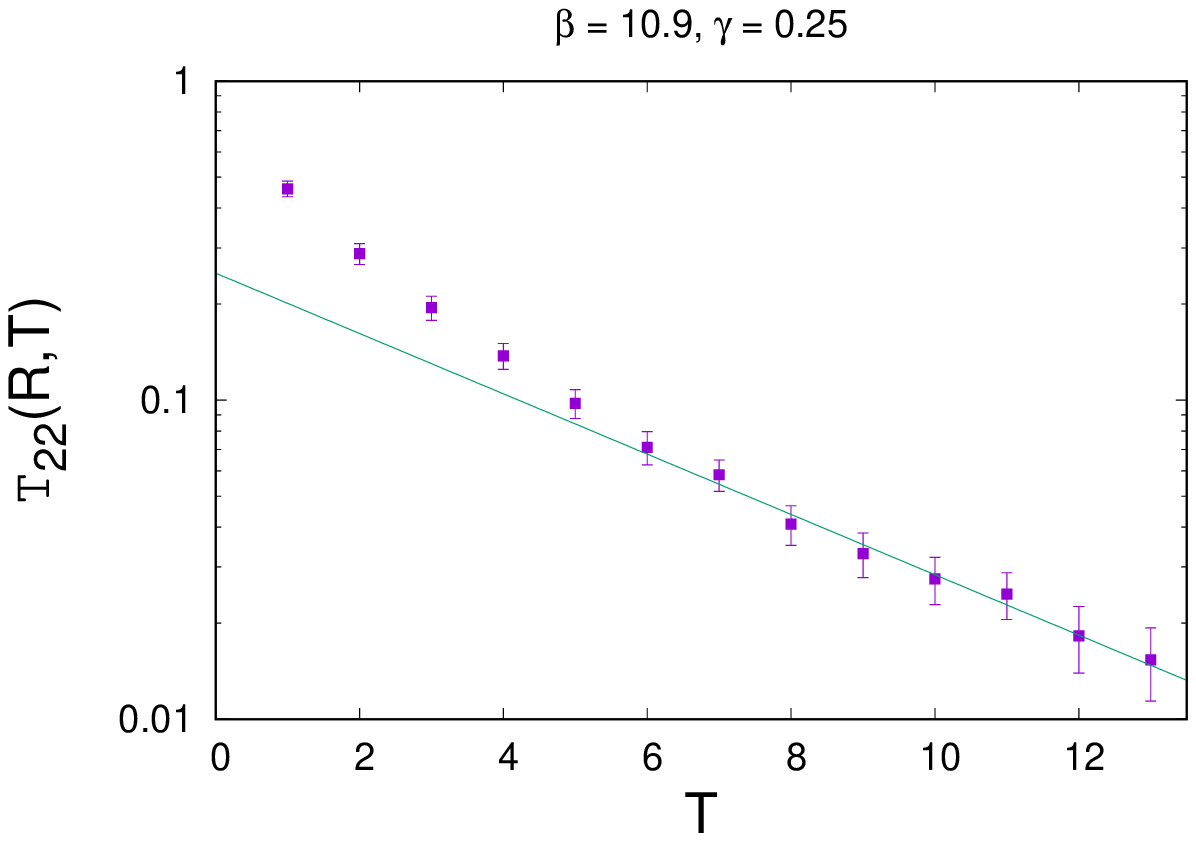} 
\caption{}
\label{fig7b}
\end{subfigure}   
\caption{(a) An exponential fitting example of the matrix element of $\T_{11}$ at $R=5.385$, $\gamma=0.25$. (b) An exponential fitting example of the matrix element of $\T_{22}$ at $R=5.385, ~ \gamma=0.25$.}
\label{fig7}
\end{figure}

To overcome this problem, we instead diagonalize $\T^{2t_0}$ in the basis $\Phi_\a$ at each separation $R$, so that we compute the transfer matrix elements $\langle \Psi_m| \T^{2t_0} | \Psi_n \rangle = \lambda_n(t_0) \delta_{mn}$
and define   $\Psi_n(t) = \T^t \Psi_n$. Consider evolving $\Psi_1$ by $t_0$ units of Euclidean time, and suppose that after this time period $\Psi_1(t_0)$ is approximately the true ground
state in the full Hilbert space.  It follows that $\Psi_{n>1}(t_0)$ is orthogonal to the ground state, because
 $\langle \Psi_m(t_0) | \Psi_n(t_0) \rangle \propto \delta_{mn}$,
and therefore, at large $T>2t_0$ 
\bea
           \T_{22}(R,T) =  \langle \Psi_2| \T^T | \Psi_2 \rangle 
                               = \langle \Psi_2(t_0)|\T^{T-2t_0}|\Psi_2(t_0) \rangle
                               \ra \text{const} \times e^{-E_{ex} T} ~~~\text{where}~~~ E_{ex} > E_1 \ .
\eea
In  Fig.\ \ref{fig7a}, we show an example of our fitting of the transfer matrix of $\T_{11}(R,T)$ at $R=5.385$, $\gamma=0.25$. We choose $2t_0=9$, and we fit $\T_{11}$ to $f_1(T) = a_1 \exp(-b_1 T) + c_1$
We found $c_1 \ne 0$, and this means that the ground state energy $E_1 \approx 0$. Note that $b_1$ gives an excited state energy.
Then similarly, we fit the matrix element of $T_{22}(R,T)$ in the range $T >6$ to a single exponential
$f_2(T) = a_2 \exp(-b_2 T)$
as shown in Fig.\ \ref{fig7b}. The coefficient $b_2 < b_1$ gives another excitation energy.

\begin{figure}[H]
\begin{subfigure}[b]{0.5\textwidth}
\centering
\includegraphics[scale=0.5]{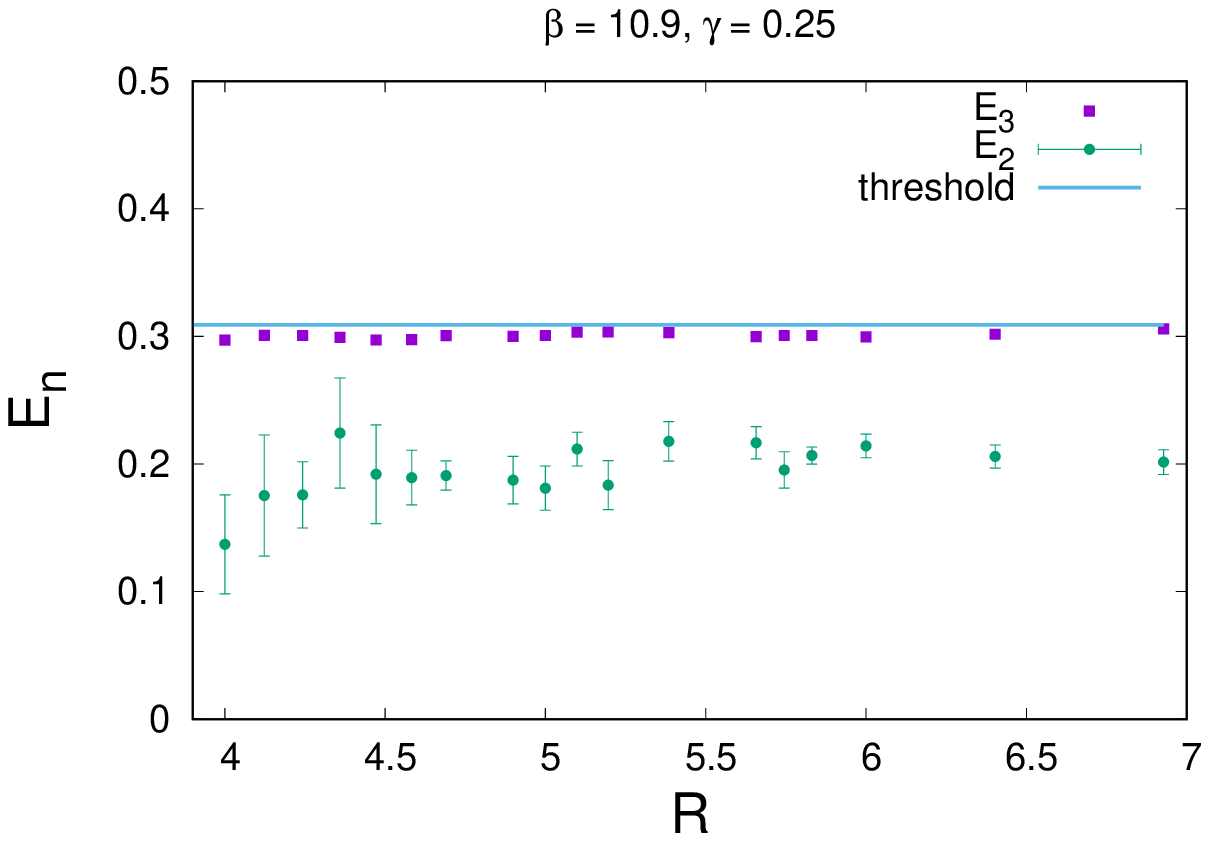} 
\caption{}
\label{fig8}
\end{subfigure}
\hfill
\begin{subfigure}[b]{0.5\textwidth}
\centering
\includegraphics[scale=0.5]{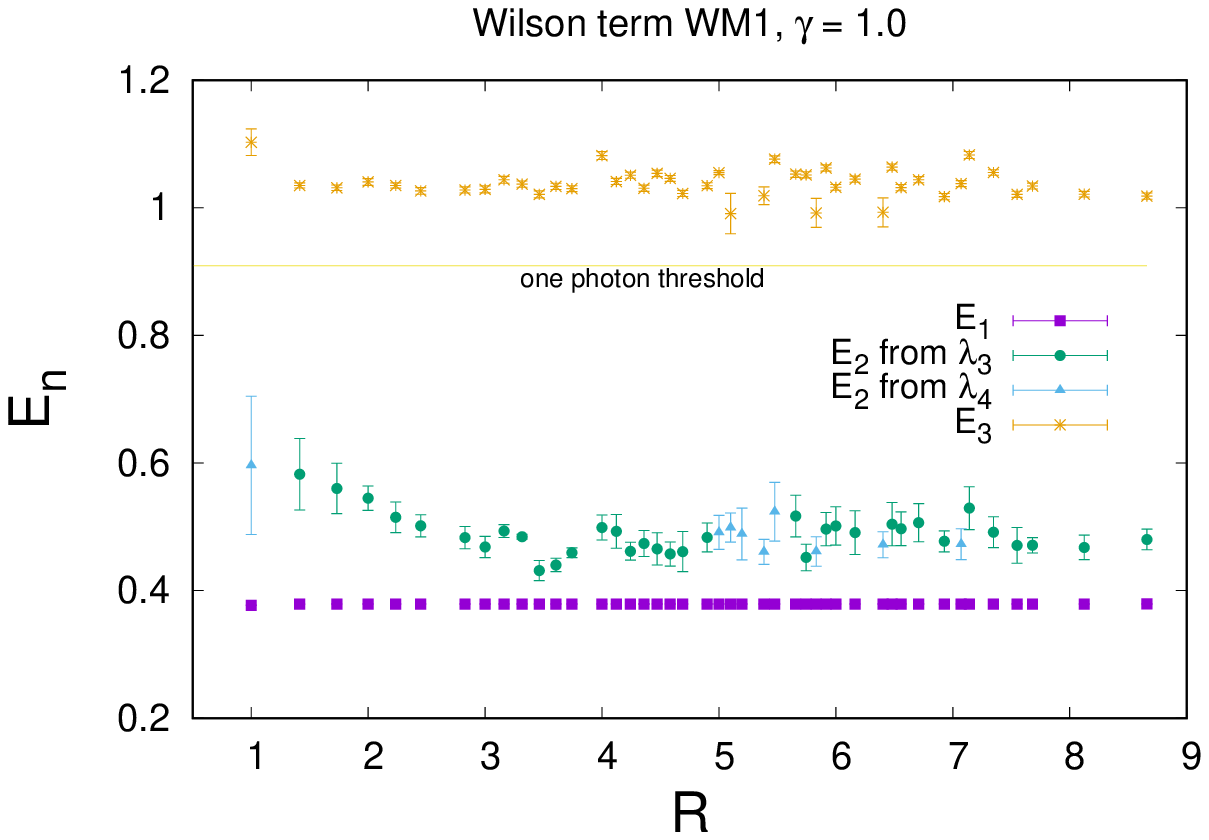} 
\caption{}
\label{fig9}
\end{subfigure}
\caption{Excitation spectrum of a static fermion and antifermion pair in (a) the effective Landau-Ginzburg model and (b) 
a chiral U(1) gauge theory (Figure from  \cite{2021_greensite}) in a Smit-Swift formulation.}  
\end{figure}

Our preliminary results (note that this is work in progress) for the excitation spectrum of the fermion and antifermion pair in effective Landau-Ginzburg model are shown in Fig.\ \ref{fig8}. In the effective Landau-Ginzburg model, we found that the data at $R<4.0$ are rather noisy, with large $\chi^2$, and these points are omitted. Note that in Fig.\ \ref{fig8} the ground state energy  of the fermion and antifermion pair is zero. 
Similarly to the previous models, we find that the first exited state of the static fermion-antifermion pair lies below the one photon threshold, at least for $R>4$.   Therefore, once again, the first excited state is stable. The second excited state, the purple dots right on the threshold in Fig.\ \ref{fig8}, is presumably the ground state plus a massive photon. 

Based on these results, we can ask if such excitations could be detected experimentally, e.g.\ by ARPES (angle-resolved photoemission spectroscopy)?  We don't yet know, but of course it would be exciting to observe such excited states in the real superconductors.

\subsection{Chiral gauge theories}
There is no known lattice formulation of chiral non-abelian gauge theories with a continuum limit. In an abelian chiral gauge theory there exists a successful formulation due to L\"{u}scher, but this formulation involves the use of overlap fermions, and it is challenging to implement numerically.

In the exploratory work by one of us \cite{2021_greensite}, a simpler option was chosen. For static fermions, work instead with a quenched version, at fixed lattice spacing, of the Smit-Swift lattice action, U(1) gauge group, with oppositely charged right and left-handed fermions. 

There are doublers, even with quenched fermions. The idea was to use a Wilson-style non-local mass term to take the mass of the doublers to infinity in the continuum. However, the continuum limit doesn't work because Smit-Swift formulation is not a true chiral gauge theory. Moreover, the positivity of the transfer matrix is unproven. But at least the non-local mass term breaks the mass degeneracy with the doublers. 

In Fig.\ \ref{fig9}, we present the numerical results for the excitation spectrum of static fermion and antifermion pair. The plot shows 
excitation energies all together $E_1,E_2,E_3$ vs.\ $R$ at $\beta=3, \gamma=1$, together with the one photon threshold. The first excited state energies are well below the one photon threshold line, and this indicates that the first excited state of the static fermion and antifermion pair is stable.  The energies of the second excited state are above the one photon threshold line, so the second excited states are probably the combination of the ground state and massive photons. Once again, our investigation in chiral gauge theory leads to the similar results of those other models of SU(3) gauge Higgs model, $q=2$ Abelian gauge Higgs model, and Landau-Ginzburg model.

\section{Conclusion}
In this work, we have shown that the gauge plus Higgs fields surrounding a charged static fermion have a spectrum of localized excitations, and these cannot be interpreted as just the ground state plus some propagating massive bosons. This means that charged ``elementary'' particles can have a mass spectrum in gauge Higgs theories. This conclusion seems robust because we see those excitation spectrums in four different models of SU(3) gauge Higgs, q=2 Abelian Higgs, Landau-Ginzburg, and chiral U(1) gauge Higgs models. Perhaps it is possible to observe those localized excitations in ARPES studies, e.g.\ in core electron spectra found by ARPES studies of conventional superconductors above and below the transition temperature. Finally, we are also interested in extending our investigation to electroweak theory, and looking for similar kinds of localized excitations of quarks and leptons, and possibly also excitations of massive gauge bosons.



\paragraph{ 
Funding for this research was provided by the United States Department of Energy under Grant No.\  DE-SC0013682.}







\nolinenumbers

\end{document}